\documentclass[prl,twocolumn,aps,showpacs]{revtex4}
\usepackage{graphicx,bm,amsmath,amssymb,psfrag}
\newcommand{\Tr}[1]{\mbox{Tr}\{ #1 \}}
\newcommand{\mscr}[1]{{\mbox{\scriptsize #1}}}
\begin{document}
\title{Determining the spin Hall conductance via charge transport}
\author{Sigurdur I.\ Erlingsson}
\author{Daniel Loss}
\affiliation{Department of Physics
  and Astronomy, University of Basel, Klingelbergstrasse 82, CH-4056,
  Switzerland}
\begin{abstract}
We propose a scheme where transport measurements of charge current and
its noise can be used to determine the spin Hall conductance in a
four-terminal setup. Starting from the scattering formalism we express the
spin current and spin Hall conductance in terms of spin-dependent transmission
coefficients.    These coefficients are then expressed in terms of charge
current and noise. We use the scheme to characterize the spin injection
efficiency of a ferromagnetic/semiconductor interface.  
\end{abstract}
\pacs{72.25.Dc,72.70.+m,71.70.Ej,85.75.-d}
\maketitle
The spin Hall effect (SHE) due to spin-orbit interaction is one of many
spintronics 
\cite{awschalom02:xx,wolf01:1488} phenomena currently under
intense study.
Initially the SHE was considered as an impurity driven
effect \cite{dyakonov71:467,hirsch99:1834}, referred to as the extrinsic SHE.
More recently, theoretical work predicting an intrinsic version of the 
SHE \cite{murakami03:1348,sinova04:126603} has sparked
renewed interest in this phenomenon.
Subsequent papers have dealt with both the effects of impurities
\cite{schliemann04:165315,inoue04:041303R,mishchenko04:226602,raimondi04:033311,chalaev:0407342} and
finite size effects within the scattering formalism
\cite{hankiewicz04:241301R,sheng04:16602,nikolic:0408693,li:0502102} on the
intrinsic SHE.
Recently the extrinsic SHE
was observed in electron doped systems \cite{kato04:1910}, where
the spin accumulation at the sample edges was detected via Kerr rotation
and from this the spin current was inferred \cite{kato04:1910}.  
Also, observation of  the intrinsic SHE in two dimensional
hole systems  has been reported \cite{wunderlich05:047204}.

The characterization of spin dependent transport via current-current
correlations (noise) has received considerable 
attention recently as well.  The charge current noise was proposed as a 
means to determine spin entanglement \cite{burkard00:16303R,egues02:176401},
and the full counting statistics of spin entangled electrons was considered
\cite{taddei02:075317}.   
On the other hand, the spin current noise was proposed as a way of probing
interactions \cite{sauret04:106601} and the full
counting statistics of spin currents were investigated
\cite{diLorenzo04:046601}.  

In this letter we propose a method of using {\em charge} current and {\em
  charge} current noise to determine the spin current and, in particular, the 
  spin Hall conductance in a four-terminal  setup.  
Our proposed scheme is quite flexible and can be applied to many different
systems as long as 
they can be described by the scattering formalism.
Quite remarkably, the scheme does not require the detection of individual spin
components of the current, and both the spin currents and spin conductances
can be expressed in terms of electrically detected quantities, i.e.\  
the charge current and noise.

{\em Spin Hall effect}.  
In extended systems with spin-orbit coupling the spin 
current density is defined as
$j_\gamma^\eta(\bm{r})=\frac{1}{2}\{j_\gamma(\bm{r}),\hat{\sigma}_\eta\}$,
where $j_\gamma(\bm{r})$ is the particle current density operator and
$\hat{\sigma}_\eta$ is a Pauli matrix
\cite{murakami03:1348,sinova04:126603,schliemann04:165315,inoue04:041303R,mishchenko04:226602,raimondi04:033311,chalaev:0407342}.  
We note that this spin current does not fulfill a continuity equation
in systems with spin-orbit interaction
\cite{rashba03:241315R,burkov05:155308,erlingsson05:035319} and its connection
to spin accumulation is still not clarified.

\begin{figure}[t]
\begin{center}
\includegraphics[angle=0,width=5cm]{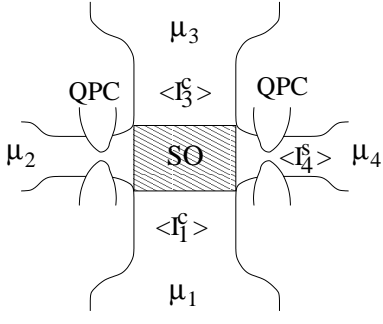}
\end{center}
\caption{Spin Hall effect in a four-terminal setup where a charge current
  flows from lead 1 to   3, resulting in spin currents in leads 2 and 4.  No
  charge current flows in lead 4, see text for details.
  Only the hatched central region has spin-orbit (SO) coupling and is
  connected to leads 2 and 4 by single mode QPCs.  The chemical potentials are
  $\mu_i=\mu_0+eV_i$.}  
\label{fig:fig1}
\end{figure}

We consider a four-terminal device where a
charge current is driven by an applied voltage
between two leads, e.g.\ lead 1 and 3 in the setup
depicted in Fig.\ \ref{fig:fig1}, and at the same time measuring the
voltage, or current, which develops in the two transverse leads 2 and 4.  
Spin-orbit interaction in the sample 
plays the role of
the magnetic field in the standard Hall setup and gives rise to a
spin current flowing perpendicularly to the charge
current driven by the applied voltage
\cite{dyakonov71:467,hirsch99:1834,murakami03:1348,sinova04:126603,schliemann04:165315,inoue04:041303R,mishchenko04:226602,raimondi04:033311,chalaev:0407342,rashba03:241315R,burkov05:155308,erlingsson05:035319}.
The leads are
assumed to have no spin-orbit interactions
\cite{hankiewicz04:241301R,sheng04:16602,nikolic:0408693,li:0502102} 
and thus the spin current is well defined.
In the scattering formalism \cite{blanter00:1} the spin current then becomes 
\begin{equation}
\langle I_{\alpha}^{s} \rangle=\frac{e}{h}\sum_\beta
\int dE \,\Tr{ \hat{\sigma}_z
  \hat{s}_{\alpha\beta}\hat{s}_{\alpha\beta}^\dagger} (f_\alpha -f_\beta), 
\label{eq:spinCurrent}
\end{equation}
where $\hat{s}_{\alpha\beta}$ are scattering matrix elements between leads
$\alpha$ and $\beta$, and $f_\alpha=f(E-\mu_\alpha)$ is the Fermi function in
lead $\alpha$ at chemical potential $\mu_\alpha$. 
The charge current $\langle I_\alpha^c \rangle $ in lead $\alpha$ is given by a
relation similar to Eq.\ (\ref{eq:spinCurrent}), with
$\hat{\sigma}_z$ replaced by the identity.  
The energy argument of the Fermi function has been suppressed,
similarly for 
the scattering matrices $\hat{s}_{\alpha\beta}$ which in
general also depend on energy. 
The hat denotes the spin structure of the scattering matrix
\begin{equation}
\hat{s}_{\alpha\beta}=
\left (
\begin{array}{cc}
s_{\alpha\beta}^{\uparrow\uparrow} & s_{\alpha\beta}^{\uparrow\downarrow} \\
s_{\alpha\beta}^{\downarrow\uparrow} & s_{\alpha\beta}^{\downarrow\downarrow} 
\end{array}
\label{eq:scatteringMatrixElement}
\right ),
\end{equation}
where $s_{\alpha\beta}^{ss'}$ denotes scattering of a transverse mode in lead
$\beta$ with spin $s'$ to a transverse mode in lead $\alpha$ with spin $s$.
Note that each element $s_{\alpha\beta}^{ss'}$ in 
Eq.\ (\ref{eq:scatteringMatrixElement}) is still a $N_\alpha\times N_\beta$
matrix in transverse mode space.
If lead $\alpha$ has only one single transverse mode, the matrix product in
Eq.\ (\ref{eq:spinCurrent}) becomes 
\begin{equation}
\hat{s}_{\alpha\beta} \hat{s}_{\alpha\beta}^\dagger 
=\left ( \begin{array}{cc}
T_{\alpha\beta}^{\uparrow\uparrow} & T_{\alpha\beta}^{\uparrow\downarrow} \\ 
(T_{\alpha\beta}^{\uparrow\downarrow})^* &
T_{\alpha\beta}^{\downarrow\downarrow}  
\end{array}
\label{eq:transmissionMatrix}
\right ),
\end{equation}
where the transmission coefficients $T_{\alpha\beta}^{ss'}$
are numbers, possibly complex for the
off-diagonal ones, irrespective of the number of transverse modes in lead
$\beta$. 
As long as the spin quantization axis of all leads is chosen to be the same
the off-diagonal matrix elements in Eq.\ (\ref{eq:transmissionMatrix}) are
  real \cite{brataas00:2481}.
The spin and charge current in lead $\alpha$ is thus
\begin{equation}
\langle I_{\alpha}^{s/c} \rangle=\frac{e}{h} \sum_\beta
\int dE (T_{\alpha\beta}^{\uparrow\uparrow} \mp
T_{\alpha\beta}^{\downarrow\downarrow})(f_\alpha - f_\beta).
\label{eq:spinCurrentT_ab}
\end{equation}

In the following we consider the linear transport regime, where the
conductances are given by spin dependent transmission coefficients, similar to 
the ones in Eq.\ (\ref{eq:transmissionMatrix}).  Knowing these allows a
complete 
determination of the charge and spin currents.
To illustrate this we consider the SHE in a four-terminal setup,
see Fig.\ \ref{fig:fig1}.
We recall that in the infinite system, the (electric field driven) charge
current induces a 
pure spin current (with no accompanying charge current) flowing in
perpendicular direction
\cite{dyakonov71:467,hirsch99:1834,murakami03:1348,sinova04:126603,schliemann04:165315,inoue04:041303R,mishchenko04:226602,raimondi04:033311,chalaev:0407342,rashba03:241315R,burkov05:155308,erlingsson05:035319}. 
To achieve 
such a situation also for the setup in Fig.\ \ref{fig:fig1}, a positive
voltage $V_1$ is applied to lead 1 and a negative one $V_3$ to lead 3.  This
results in a charge current  
flowing from lead 1 to 3 and 
the values of  
$V_1$ and $V_3$ are set by the requirement of zero transverse charge current
in lead 4:
\begin{equation}
\langle I_4^c \rangle =\frac{e^2}{h} \sum_{p=\pm}\bigl (T_{41;p}
(V_4-V_1)+T_{43;p} 
(V_4-V_3) \bigr )=0.
\label{eq:zeroCurrent4}
\end{equation}
$\langle I_2^c \rangle$ is left unspecified, see also below Eq.\
(\ref{eq:spinHallConductance}). 
Here we have introduced the transmission eigenvalues $T_{41;p}$ and $T_{43;p}$
of Eq.\ (\ref{eq:transmissionMatrix}) for $\beta=1$ and $3$, {\em resp.}  
In terms of these
eigenvalues both the current and noise will take on a
simple form \cite{martin92:1742}.
The quantum point contacts (QPC) are used to ensure that only 
a single transverse mode 
couples the scattering region to leads 2 and 4.
Taking into account the constraint imposed by Eq.\
(\ref{eq:zeroCurrent4}), the spin current flowing in lead 4 takes the form
\begin{equation}
\langle I_{4}^s \rangle=
\frac{e^2}{h} 
\left (
(T_{41}^{\uparrow\uparrow}-T_{41}^{\downarrow\downarrow})
-\frac{g_{41}}{g_{43}}
(T_{43}^{\uparrow\uparrow}-T_{43}^{\downarrow\downarrow})
\right )(V_4-V_1),
\label{eq:spinCurrent4}
\end{equation}
where $g_{\alpha\beta}=\sum_p T_{\alpha\beta;p}\geq 0$ is the dimensionless
conductance between leads $\alpha$ and $\beta$.
To proceed further we make the following assumptions about the spin-orbit
interaction.  The spin state labeled 'up' is predominantly scattered to the
right, with respect to its propagation direction. 
This results in $(T_{41}^{\uparrow\uparrow}-T_{41}^{\downarrow\downarrow})>0$
and $(T_{43}^{\uparrow\uparrow}-T_{43}^{\downarrow\downarrow})<0$, since
electrons going from lead 3 to 4 'turn left' with respect to their propagation
direction.  This assumption
is not very restrictive and is satisfied by various known spin-orbit scattering
mechanisms
\cite{dyakonov71:467,hirsch99:1834,murakami03:1348,sinova04:126603,schliemann04:165315,inoue04:041303R,mishchenko04:226602,raimondi04:033311,chalaev:0407342,rashba03:241315R,burkov05:155308,erlingsson05:035319,mott29:425}.
With this we write the spin current as
\begin{equation}
\langle I_{4}^s \rangle=
\frac{e^2}{h} 
\left (
|T_{41}^{\uparrow\uparrow}-T_{41}^{\downarrow\downarrow}|
+\frac{g_{41}}{g_{43}}
|T_{43}^{\uparrow\uparrow}-T_{43}^{\downarrow\downarrow}|
\right )(V_4-V_1).
\label{eq:spinCurrent4abs}
\end{equation}
Unless one knows the details of the spin-orbit scattering one cannot tell
whether spin $\uparrow$ or $\downarrow$ is scattered more to the right and so
the direction (sign) of the spin current is unknown.  
From the spin current in Eq.\ (\ref{eq:spinCurrent4abs}) we define 
the spin Hall conductance as ($e>0$)
\begin{eqnarray}
G_H^s&=&\frac{\frac{\hbar}{2e}\langle I_4^s\rangle }{V_1-V_3} \\
&=&\frac{e}{4\pi} 
\frac{
\left (
|T_{41}^{\uparrow\uparrow}-T_{41}^{\downarrow\downarrow}|
+\frac{g_{41}}{g_{43}}
|T_{43}^{\uparrow\uparrow}-T_{43}^{\downarrow\downarrow}|
\right )
}
{(1+\frac{g_{41}}{g_{43}})}.
\label{eq:spinHallConductance}
\end{eqnarray}
Eq.\ (\ref{eq:spinHallConductance}) gives the spin Hall conductance
$G_H^s$, up to 
a sign, as a function of the spin dependent transmission
coefficients. 
From this we also see that the maximal possible spin Hall conductance is
$G^s_H=e/4\pi$, this bound resulting from the QPC's single transverse mode. 
As we will show below, the spin dependent transmission coefficients
$T_{\alpha\beta}^{ss'}$ can be determined using standard transport
measurements of charge current and noise.  
In Eq.\ (\ref{eq:spinHallConductance}) no assumption  was made about the sample
symmetry.
In general it is not possible to simultaneously
ensure $\langle I_2^c\rangle=0$ and $\langle I_4^c \rangle=0$, due to
asymmetries.  However, this is not critical since $G_H^s$ is defined using
only 
one of the two transverse leads.
For symmetric structures
$g_{43}=g_{41}$ and
$|T_{41}^{\uparrow\uparrow}-T_{41}^{\downarrow\downarrow}|=|T_{43}^{\uparrow\uparrow}-T_{43}^{\downarrow\downarrow}|$,
which reduces the number of measurements required to determine
$T_{\alpha\beta}^{ss'}$.
This symmetry will be broken if a perpendicular magnetic field is applied
\cite{shepard92:9648}.

Let us now determine $T_{41}^{ss'}$.  
With all leads at chemical potential $\mu_0$, except for lead 1 where
a bias $V_1$ is applied, the charge current in lead 4 
is written in terms of the transmission eigenvalues as 
$\langle I_4^c \rangle=\frac{e^2V_1}{h}\sum_{p=\pm} T_{41;p}$.

Next we introduce the excess noise $\Delta S_{\alpha\beta}$, which is the total
noise $S_{\alpha\beta}$  
minus the Johnson-Nyquist contribution
$2\frac{k_BTe^2}{h}(g_{\alpha\beta}+g_{\beta\alpha})$ \cite{blanter00:1}.
The excess noise in lead 4 becomes (voltage $V_1$ applied to lead 1)
\begin{eqnarray}
\Delta S_{44}&=&
\frac{2e^2}{h}\int dE
\Tr{\hat{s}_{41}\hat{s}_{41}^\dagger(1-\hat{s}_{41}\hat{s}_{41}^\dagger)}
(f_1-f_0)^2
\nonumber \\
&=&\frac{2e^3}{h} V_1 F\left( \frac{eV_1}{k_BT} \right)
\sum_{p=\pm} T_{41;p}(1- T_{41;p}),
\label{eq:noise44_1}
\end{eqnarray}
where $f_0=f(E-\mu_0)$ and  $F(x)=\coth(x/2)-2/x$ \cite{lesovik89:592}.  
In the following we will assume that the transmission eigenvalues do
not change much 
\footnote[1]{
Applying the voltage symmetrically around $\mu_0$ the component of the 
 transmission coefficent which is linear in energy vanishes upon integration,
 and deviations $\delta T_{\alpha\beta}^{ss'}(E)$ are at 
 least quadratic in energy away from $\mu_0$.
  To resolve the spin dependence of the transmission coefficients we require
 that $\delta T_{\alpha\beta}^{ss'}|_\mscr{max} \ll |
 T_{\alpha\beta}^{\uparrow\uparrow} -T_{\alpha\beta}^{\downarrow\downarrow}|$,
  where $\delta T_{\alpha\beta}^{ss'}|_\mscr{max}$ is the maximum deviation
 within the bias window. 
}
within the bias window, but apart from that only the unitarity
of the scattering matrix was used to derive Eq.\ (\ref{eq:noise44_1}).
From the charge current $\langle I_4^c \rangle$
and Eq.\ (\ref{eq:noise44_1}) we obtain 
\begin{eqnarray}
T_{41;\pm}=\frac{1}{2}g_{41}\pm
\frac{1}{2}
\sqrt{g_{41}(2-g_{41})-\frac{h\Delta S_{44}}{e^3|V_1|}}.
\label{eq:transmissionEigenvalues41}
\end{eqnarray}
The conductance is determined by the current, 
$g_{41}=\frac{h\langle I_4^c \rangle}{e^2V_1}$.
For simplicity we have assumed $e|V_1|/k_BT\gg 1$ but the results are easily
generalized to arbitrary temperatures by 
$|eV_1|\rightarrow eV_1 F(eV_1/k_B T)$.

In terms of $\delta T_{41}=T_{41;+}-T_{41;-}$ the excess noise in Eq.\
(\ref{eq:noise44_1}) can be written as $\Delta S_{44}=
\frac{e^3 |V_1|}{h} (g_{41}(2-g_{41})-\delta T_{41}^2 )>0$.
This
relation shows that the excess noise $\Delta S_{44}$ decreases with increasing
$\delta T_{41}$, ensuring that the argument of 
the square root in Eq.\ (\ref{eq:transmissionEigenvalues41}) is positive.

According to Eq.\ (\ref{eq:spinHallConductance}) the spin Hall conductance is
in part determined by
$|T_{41}^{\uparrow\uparrow}-T_{41}^{\downarrow\downarrow}|$. 
This quantity can be written in terms of the
transmission eigenvalues as
\begin{equation}
|T_{41}^{\uparrow\uparrow}-T_{41}^{\downarrow\downarrow}|=
\sqrt{(T_{41;+}-T_{41;-})^2-4|T_{41}^{\uparrow\downarrow}|^2}.
\label{eq:transmissionCoefficientDifference}
\end{equation}
If the off-diagonal terms $T_{\alpha\beta}^{ss'}$ vanish then
the spin current Eq.\ (\ref{eq:spinCurrent4abs}) is completely determined by the
transmission eigenvalues, i.e.\
$|T_{41}^{\uparrow\uparrow}-T_{41}^{\downarrow\downarrow}|
=|T_{41;+}-T_{41;-}|$.  
When $|T_{41}^{\uparrow\downarrow}|\neq 0$, additional measurements are
necessary.  
With a voltage  $V_2$  applied to lead 2
the charge current and excess noise in lead 4 are given by 
\begin{eqnarray}
\langle I_4^c \rangle&=&
\frac{e^2V_2}{h}
\sum_{p=\pm} T_{42;p}\,\,,
\label{eq:current42} \\ 
\Delta S_{44}&=&
\frac{2e^3|V_2|}{h}
\sum_{p=\pm} T_{42;p}(1- T_{42;p}).
\label{eq:noise42}
\end{eqnarray}
For non-magnetic leads and forward scattering (from lead 4 to 2) we have
$T_{42}^{\uparrow\uparrow}=T_{42}^{\downarrow\downarrow}$
\cite{molenkamp01:121202R}, which results in
$T_{42;\pm}=T_{42}^{\uparrow\uparrow}\pm 
|T_{42}^{\uparrow\downarrow}|$.  Assuming that $T_{42}^{\uparrow\downarrow}$
is real we get
\begin{equation}
|T_{42}^{\uparrow\downarrow}|
=\frac{1}{2}\sqrt{g_{42}(2-g_{42})-\frac{h\Delta S_{44}}{e^3|V_2|}}.
\end{equation}

Applying a voltage $V_4$ to lead 4, the
charge current cross-correlation between leads 2 and 1 gives
$T_{41}^{\uparrow\downarrow}$ via the relation 
\begin{eqnarray}
\Delta S_{21}&=&
-\frac{2e^2}{h}\int dE \Tr{\hat{s}_{42}\hat{s}_{42}^\dagger
  \hat{s}_{41}\hat{s}_{41}^\dagger} (f_4-f_0)^2 \nonumber \\
&=&-\frac{2e^3|V_4|}{h}
\bigl (\frac{1}{2}g_{42} g_{41} +2T_{42}^{\uparrow\downarrow}
T_{41}^{\uparrow \downarrow} \bigr)  .
\label{eq:noise41_4}
\end{eqnarray}
In deriving Eq.\ (\ref{eq:noise41_4}) time reversal symmetry was assumed,
i.e.\ in the absence of a magnetic field $B$.  
[We note in passing that this can be generalized to 
$B\neq 0$ by incorporating the appropriate symmetries of the 
scattering matrix $\hat{s}_{\alpha\beta}(+B)=(i\Sigma_{y;\alpha})^{-1}  
(\hat{s}_{\beta\alpha}^\dagger(-B))^*(i\Sigma_{y;\beta})
$ where $i\Sigma_{y;\alpha}=
\left ( 
\begin{array}{cc}
0 &\mathbb{I}_\alpha \\
-\mathbb{I}_\alpha & 0 
\end{array} 
\right )
$ comes from the time reversal operation.]
The off-diagonal element can thus be expressed in terms
of charge transport quantities
\begin{eqnarray}
4|T_{41}^{\uparrow\downarrow}|^2&=&
\frac{
\left (
\frac{h \Delta S_{21}(V_4)}{e^3|V_4|}+g_{41}g_{42}
\right )^2
}{
g_{42}(2-g_{42})-\frac{h\Delta S_{44}(V_2)}{e^3| V_2|}
}.
\end{eqnarray}
Here we have added an argument to each noise measurement to indicate that,
e.g., \ $\Delta S_{44}(V_2)$ corresponds  to the excess noise in lead 4 with a
finite voltage $V_2$ applied only to lead 2, again assuming
$T_{\alpha\beta}^{ss'}$ to be constant \footnotemark[34].

With the same procedure we obtain $T_{43}^{ss'}$:  $(i)$ apply
voltage 
$V_3$ to lead 3 and measure $\Delta 
S_{44}$, which gives $T_{43;\pm}$, $(ii)$ apply voltage $V_4$ to lead 4 and
measure $\Delta S_{23}$, which gives
$|T_{43}^{\uparrow\downarrow}|$.  All these measured quantities taken together
yield the spin Hall conductance from Eq.\ (\ref{eq:spinHallConductance})
\begin{eqnarray}
G_H^s&=&
\frac{e}{4\pi}\sum_{\beta=1,3}
\frac{g_{4\beta}^{-1}}{g_{41}^{-1}+g_{43}^{-1}}
\Biggl \{
g_{4\beta}(2-g_{4\beta})
\nonumber \\
&-&
\frac{h\Delta S_{44}(V_\beta)}{e^3| V_\beta|} 
-\frac{ \left (
\frac{h \Delta S_{2\beta}(V_4)}{e^3|V_4|}+g_{4\beta}g_{42}
\right )^2
}{
g_{42}(2-g_{42})-\frac{h\Delta S_{44}(V_2)}{e^3| V_2|}
}
\Biggr \}^\frac{1}{2} 
\!\!\!\!\!.
\end{eqnarray}
This equation is the central result of our paper from which we see that the
spin Hall conductance $G_H^s$ can be solely obtained by charge transport
measurements without the need of any spin dependent detection. 
This result applies equally to the extrinsic and intrinsic spin Hall effect,
with no size restriction on the SO scattering region.
Our scheme is restricted to only a
single transverse mode in the QPCs.  
In case of more transverse modes, but still few, numerical analysis can
be used to extract information about the transmission eigenvalues
\cite{vandenBrom99:1526}.
However, since the spin polarization only comes from the
'active region' we expect that opening up the QPC (for larger current)
should not greatly change the amount of spin polarization determined for
the single mode case. 

\begin{figure}[h]
\begin{center}
\includegraphics[angle=0,width=5cm]{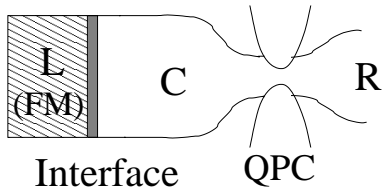}
\end{center}
\caption{A schematic setup of spin injection from a magnetic (FM)
   lead $L$ into a semiconducting central region lead $C$ which is connected
   to lead $R$ via the QPC. The interface between $L$ and $C$ can be clean or
   disordered.}  
\label{fig:fig2}
\end{figure}

{\em Spin injection}.  Here we consider a hetero-junction,
consisting of a semiconductor and a magnetic material (ferromagnet or
magnetic semiconductor) which serves as an spin injector, see Fig.\
\ref{fig:fig2}. 
This setup allows the characterization of the spin injection from lead $L$ to
lead $C$.  
The QPC, which is assumed to have known scattering properties,  
ensures a single transverse mode in  the detection 
lead $R$.
The number of transverse modes in leads $C$ and $L$ is
arbitrary, so the area of the interface cross-section is not constrained.  
The spin quantization axis is chosen along the magnetization direction of the
ferromagnet and it is assumed that $T_{LR}^{\uparrow\downarrow} \ll
T_{LR}^{\uparrow\uparrow},T_{LR}^{\downarrow\downarrow}$, which is a good
approximation for ferromagnets \cite{brataas00:2481}.
In this case only a single charge current and noise measurement is needed to
obtain
$|T_{LR}^{\uparrow\uparrow}-T_{LR}^{\downarrow\downarrow}|$,
from which the polarization $p$ of the injected current is obtained
 from Eqs.\ (\ref{eq:spinCurrentT_ab}),
 (\ref{eq:transmissionEigenvalues41}) and
 (\ref{eq:transmissionCoefficientDifference})  (replacing the subscripts
 $4,1$ with $L,R$) 
\begin{equation}
p=\frac{\langle I_{R}^s \rangle}{\langle I_{R}^c \rangle}=
\frac{1}{g_{LR}}
\sqrt{g_{LR}(2-g_{LR})-\frac{h\Delta S_{RR}(V_L)}{e^3|V_L|}}, 
\label{eq:polarizationLR}
\end{equation}
where $V_L$ is the bias applied  at lead $L$.
Since the scattering from $C$ to $R$
 is assumed to be non-spin-selective, it is a reasonable assumption
 that the polarization obtained via Eq.\ (\ref{eq:polarizationLR})
 characterizes the spin polarization of the interface between leads $L$ and
 $C$, even for a QPC with many transverse modes.

In summary, we have introduced a scheme using charge current and noise
measurements to 
extract spin polarization, 
without the need of spin-resolved measurements.
The theory is based on the scattering formalism which makes this scheme quite
flexible and applicable to many different systems.

The authors would like thank D.\ Saraga, G.\ Burkard, J.\ C.\ Egues,
W.\ Belzig, J.\ Schliemann and O.\ Chalaev for helpful discussions.
This work was financed by the Spintronics RTN, the NCCR
Nanoscience, the Swiss NSF, DARPA, ARO, ONR.


\begin{thebibliography}{33}
\expandafter\ifx\csname natexlab\endcsname\relax\def\natexlab#1{#1}\fi
\expandafter\ifx\csname bibnamefont\endcsname\relax
  \def\bibnamefont#1{#1}\fi
\expandafter\ifx\csname bibfnamefont\endcsname\relax
  \def\bibfnamefont#1{#1}\fi
\expandafter\ifx\csname citenamefont\endcsname\relax
  \def\citenamefont#1{#1}\fi
\expandafter\ifx\csname url\endcsname\relax
  \def\url#1{\texttt{#1}}\fi
\expandafter\ifx\csname urlprefix\endcsname\relax\def\urlprefix{URL }\fi
\providecommand{\bibinfo}[2]{#2}
\providecommand{\eprint}[2][]{\url{#2}}

\bibitem[{\citenamefont{Awschalom et~al.}(2002)\citenamefont{Awschalom, Loss,
  and Samarth}}]{awschalom02:xx}
\bibinfo{editor}{\bibfnamefont{D.~D.} \bibnamefont{Awschalom}},
  \bibinfo{editor}{\bibfnamefont{D.}~\bibnamefont{Loss}}, \bibnamefont{and}
  \bibinfo{editor}{\bibfnamefont{N.}~\bibnamefont{Samarth}}, eds.,
  \emph{\bibinfo{title}{Semiconductor Spintronics and Quantum Computation}}
  (\bibinfo{publisher}{Springer-Verlag}, \bibinfo{address}{Berlin},
  \bibinfo{year}{2002}).

\bibitem[{\citenamefont{Wolf et~al.}(2001)\citenamefont{Wolf, Awschalom,
  Buhrman, Daughton, von Moln\'ar, Roukes, Chtchelkanova, and
  Treger}}]{wolf01:1488}
\bibinfo{author}{\bibfnamefont{S.~A.} \bibnamefont{Wolf}},
  \bibinfo{author}{\bibfnamefont{D.~D.} \bibnamefont{Awschalom}},
  \bibinfo{author}{\bibfnamefont{R.~A.} \bibnamefont{Buhrman}},
  \bibinfo{author}{\bibfnamefont{J.~M.} \bibnamefont{Daughton}},
  \bibinfo{author}{\bibfnamefont{S.}~\bibnamefont{von Moln\'ar}},
  \bibinfo{author}{\bibfnamefont{M.~L.} \bibnamefont{Roukes}},
  \bibinfo{author}{\bibfnamefont{A.~Y.} \bibnamefont{Chtchelkanova}},
  \bibnamefont{and} \bibinfo{author}{\bibfnamefont{D.~M.}
  \bibnamefont{Treger}}, \bibinfo{journal}{Science}
  \textbf{\bibinfo{volume}{294}}, \bibinfo{pages}{1488} (\bibinfo{year}{2001}).

\bibitem[{\citenamefont{Dyakonov and Perel}(1971)}]{dyakonov71:467}
\bibinfo{author}{\bibfnamefont{M.~I.} \bibnamefont{Dyakonov}} \bibnamefont{and}
  \bibinfo{author}{\bibfnamefont{V.~I.} \bibnamefont{Perel}},
  \bibinfo{journal}{JETP Lett.} \textbf{\bibinfo{volume}{13}},
  \bibinfo{pages}{467} (\bibinfo{year}{1971}).

\bibitem[{\citenamefont{Hirsch}(1999)}]{hirsch99:1834}
\bibinfo{author}{\bibfnamefont{J.~E.} \bibnamefont{Hirsch}},
  \bibinfo{journal}{Phys.\ Rev.\ Lett.} \textbf{\bibinfo{volume}{83}},
  \bibinfo{pages}{1834} (\bibinfo{year}{1999}).

\bibitem[{\citenamefont{Murakami et~al.}(2003)\citenamefont{Murakami, Nagaosa,
  and Zhang}}]{murakami03:1348}
\bibinfo{author}{\bibfnamefont{S.}~\bibnamefont{Murakami}},
  \bibinfo{author}{\bibfnamefont{N.}~\bibnamefont{Nagaosa}}, \bibnamefont{and}
  \bibinfo{author}{\bibfnamefont{S.-C.} \bibnamefont{Zhang}},
  \bibinfo{journal}{Science} \textbf{\bibinfo{volume}{301}},
  \bibinfo{pages}{1348} (\bibinfo{year}{2003}).

\bibitem[{\citenamefont{Sinova et~al.}(2004)\citenamefont{Sinova, Culcer, Niu,
  Sinitsyn, Jungwirth, and MacDonald}}]{sinova04:126603}
\bibinfo{author}{\bibfnamefont{J.}~\bibnamefont{Sinova}},
  \bibinfo{author}{\bibfnamefont{D.}~\bibnamefont{Culcer}},
  \bibinfo{author}{\bibfnamefont{Q.}~\bibnamefont{Niu}},
  \bibinfo{author}{\bibfnamefont{N.~A.} \bibnamefont{Sinitsyn}},
  \bibinfo{author}{\bibfnamefont{T.}~\bibnamefont{Jungwirth}},
  \bibnamefont{and} \bibinfo{author}{\bibfnamefont{A.~H.}
  \bibnamefont{MacDonald}}, \bibinfo{journal}{Phys.\ Rev.\ Lett.}
  \textbf{\bibinfo{volume}{92}}, \bibinfo{pages}{126603}
  (\bibinfo{year}{2004}).

\bibitem[{\citenamefont{Schliemann and Loss}(2004)}]{schliemann04:165315}
\bibinfo{author}{\bibfnamefont{J.}~\bibnamefont{Schliemann}} \bibnamefont{and}
  \bibinfo{author}{\bibfnamefont{D.}~\bibnamefont{Loss}},
  \bibinfo{journal}{Phys.\ Rev.\ B} \textbf{\bibinfo{volume}{69}},
  \bibinfo{pages}{165315} (\bibinfo{year}{2004}).

\bibitem[{\citenamefont{Inoue et~al.}(2004)\citenamefont{Inoue, Bauer, and
  Molenkamp}}]{inoue04:041303R}
\bibinfo{author}{\bibfnamefont{J.}~\bibnamefont{Inoue}},
  \bibinfo{author}{\bibfnamefont{G.~E.~W.} \bibnamefont{Bauer}},
  \bibnamefont{and} \bibinfo{author}{\bibfnamefont{L.~W.}
  \bibnamefont{Molenkamp}}, \bibinfo{journal}{Phys.\ Rev.\ B}
  \textbf{\bibinfo{volume}{70}}, \bibinfo{pages}{041303R}
  (\bibinfo{year}{2004}).

\bibitem[{\citenamefont{Mishchenko et~al.}(2004)\citenamefont{Mishchenko,
  Shytov, and Halperin}}]{mishchenko04:226602}
\bibinfo{author}{\bibfnamefont{E.~G.} \bibnamefont{Mishchenko}},
  \bibinfo{author}{\bibfnamefont{A.~V.} \bibnamefont{Shytov}},
  \bibnamefont{and} \bibinfo{author}{\bibfnamefont{B.~I.}
  \bibnamefont{Halperin}}, \bibinfo{journal}{Phys.\ Rev.\ Lett.}
  \textbf{\bibinfo{volume}{93}}, \bibinfo{pages}{226602}
  (\bibinfo{year}{2004}).

\bibitem[{\citenamefont{Raimondi and Schwab}(2005)}]{raimondi04:033311}
\bibinfo{author}{\bibfnamefont{R.}~\bibnamefont{Raimondi}} \bibnamefont{and}
  \bibinfo{author}{\bibfnamefont{P.}~\bibnamefont{Schwab}},
  \bibinfo{journal}{Phys.\ Rev.\ B} \textbf{\bibinfo{volume}{71}},
  \bibinfo{pages}{033111} (\bibinfo{year}{2005}).

\bibitem[{\citenamefont{Chalaev and Loss}()}]{chalaev:0407342}
\bibinfo{author}{\bibfnamefont{O.}~\bibnamefont{Chalaev}} \bibnamefont{and}
  \bibinfo{author}{\bibfnamefont{D.}~\bibnamefont{Loss}},
  \bibinfo{note}{cond-mat/0407342 (unpublished)}.

\bibitem[{\citenamefont{Hankiewicz et~al.}(2004)\citenamefont{Hankiewicz,
  Molenkamp, Jungwirth, and Sinova}}]{hankiewicz04:241301R}
\bibinfo{author}{\bibfnamefont{E.~M.} \bibnamefont{Hankiewicz}},
  \bibinfo{author}{\bibfnamefont{L.~W.} \bibnamefont{Molenkamp}},
  \bibinfo{author}{\bibfnamefont{T.}~\bibnamefont{Jungwirth}},
  \bibnamefont{and} \bibinfo{author}{\bibfnamefont{J.}~\bibnamefont{Sinova}},
  \bibinfo{journal}{Phys.\ Rev.\ B} \textbf{\bibinfo{volume}{70}},
  \bibinfo{pages}{241301R} (\bibinfo{year}{2004}).

\bibitem[{\citenamefont{Sheng et~al.}(2004)\citenamefont{Sheng, Sheng, and
  Ting}}]{sheng04:16602}
\bibinfo{author}{\bibfnamefont{L.}~\bibnamefont{Sheng}},
  \bibinfo{author}{\bibfnamefont{D.~N.} \bibnamefont{Sheng}}, \bibnamefont{and}
  \bibinfo{author}{\bibfnamefont{C.~S.} \bibnamefont{Ting}},
  \bibinfo{journal}{Phys.\ Rev.\ Lett.} \textbf{\bibinfo{volume}{94}},
  \bibinfo{pages}{016602} (\bibinfo{year}{2004}).

\bibitem[{\citenamefont{Nikolic et~al.}()\citenamefont{Nikolic, Zarbo, and
  Souma}}]{nikolic:0408693}
\bibinfo{author}{\bibfnamefont{B.~K.} \bibnamefont{Nikolic}},
  \bibinfo{author}{\bibfnamefont{L.~P.} \bibnamefont{Zarbo}}, \bibnamefont{and}
  \bibinfo{author}{\bibfnamefont{S.}~\bibnamefont{Souma}},
  \bibinfo{note}{cond-mat/0408693 (unpublished)}.

\bibitem[{\citenamefont{Li et~al.}()\citenamefont{Li, Hu, and
  Shen}}]{li:0502102}
\bibinfo{author}{\bibfnamefont{J.}~\bibnamefont{Li}},
  \bibinfo{author}{\bibfnamefont{L.}~\bibnamefont{Hu}}, \bibnamefont{and}
  \bibinfo{author}{\bibfnamefont{S.-Q.} \bibnamefont{Shen}},
  \bibinfo{note}{con-mat/0502102 (unpublished)}.

\bibitem[{\citenamefont{Kato et~al.}(2004)\citenamefont{Kato, Myers, Gossard,
  and Awschalom}}]{kato04:1910}
\bibinfo{author}{\bibfnamefont{Y.~K.} \bibnamefont{Kato}},
  \bibinfo{author}{\bibfnamefont{R.~C.} \bibnamefont{Myers}},
  \bibinfo{author}{\bibfnamefont{A.~C.} \bibnamefont{Gossard}},
  \bibnamefont{and} \bibinfo{author}{\bibfnamefont{D.~D.}
  \bibnamefont{Awschalom}}, \bibinfo{journal}{Science}
  \textbf{\bibinfo{volume}{306}}, \bibinfo{pages}{1910} (\bibinfo{year}{2004}).

\bibitem[{\citenamefont{Wunderlich et~al.}(2005)\citenamefont{Wunderlich,
  Kaestner, Sinova, and Jungwirth}}]{wunderlich05:047204}
\bibinfo{author}{\bibfnamefont{J.}~\bibnamefont{Wunderlich}},
  \bibinfo{author}{\bibfnamefont{B.}~\bibnamefont{Kaestner}},
  \bibinfo{author}{\bibfnamefont{J.}~\bibnamefont{Sinova}}, \bibnamefont{and}
  \bibinfo{author}{\bibfnamefont{T.}~\bibnamefont{Jungwirth}},
  \bibinfo{journal}{Phys.\ Rev.\ Lett.} \textbf{\bibinfo{volume}{94}},
  \bibinfo{pages}{047204} (\bibinfo{year}{2005}).

\bibitem[{\citenamefont{Burkard et~al.}(2000)\citenamefont{Burkard, Loss, and
  Sukhorukov}}]{burkard00:16303R}
\bibinfo{author}{\bibfnamefont{G.}~\bibnamefont{Burkard}},
  \bibinfo{author}{\bibfnamefont{D.}~\bibnamefont{Loss}}, \bibnamefont{and}
  \bibinfo{author}{\bibfnamefont{E.~V.} \bibnamefont{Sukhorukov}},
  \bibinfo{journal}{Phys.\ Rev.\ B} \textbf{\bibinfo{volume}{61}},
  \bibinfo{pages}{16303R} (\bibinfo{year}{2000}).

\bibitem[{\citenamefont{Egues et~al.}(2002)\citenamefont{Egues, Burkard, and
  Loss}}]{egues02:176401}
\bibinfo{author}{\bibfnamefont{J.~C.} \bibnamefont{Egues}},
  \bibinfo{author}{\bibfnamefont{G.}~\bibnamefont{Burkard}}, \bibnamefont{and}
  \bibinfo{author}{\bibfnamefont{D.}~\bibnamefont{Loss}},
  \bibinfo{journal}{Phys.\ Rev.\ Lett.} \textbf{\bibinfo{volume}{89}},
  \bibinfo{pages}{176401} (\bibinfo{year}{2002}).

\bibitem[{\citenamefont{Taddei and Fazio}(2002)}]{taddei02:075317}
\bibinfo{author}{\bibfnamefont{F.}~\bibnamefont{Taddei}} \bibnamefont{and}
  \bibinfo{author}{\bibfnamefont{R.}~\bibnamefont{Fazio}},
  \bibinfo{journal}{Phys.\ Rev.\ B} \textbf{\bibinfo{volume}{65}},
  \bibinfo{pages}{075317} (\bibinfo{year}{2002}).

\bibitem[{\citenamefont{Sauret and Feinberg}(2004)}]{sauret04:106601}
\bibinfo{author}{\bibfnamefont{O.}~\bibnamefont{Sauret}} \bibnamefont{and}
  \bibinfo{author}{\bibfnamefont{D.}~\bibnamefont{Feinberg}},
  \bibinfo{journal}{Phys.\ Rev.\ Lett.} \textbf{\bibinfo{volume}{92}},
  \bibinfo{pages}{106601} (\bibinfo{year}{2004}).

\bibitem[{\citenamefont{{Di Lorenzo} and Nazarov}(2004)}]{diLorenzo04:046601}
\bibinfo{author}{\bibfnamefont{A.}~\bibnamefont{{Di Lorenzo}}}
  \bibnamefont{and} \bibinfo{author}{\bibfnamefont{Y.~V.}
  \bibnamefont{Nazarov}}, \bibinfo{journal}{Phys.\ Rev.\ Lett.}
  \textbf{\bibinfo{volume}{93}}, \bibinfo{pages}{046601}
  (\bibinfo{year}{2004}).

\bibitem[{\citenamefont{Rashba}(2003)}]{rashba03:241315R}
\bibinfo{author}{\bibfnamefont{E.~I.} \bibnamefont{Rashba}},
  \bibinfo{journal}{Phys.\ Rev.\ B} \textbf{\bibinfo{volume}{68}},
  \bibinfo{pages}{241315R} (\bibinfo{year}{2003}).

\bibitem[{\citenamefont{Burkov et~al.}(2004)\citenamefont{Burkov, N\'u{\~n}es,
  and MacDonald}}]{burkov05:155308}
\bibinfo{author}{\bibfnamefont{A.~A.} \bibnamefont{Burkov}},
  \bibinfo{author}{\bibfnamefont{A.~S.} \bibnamefont{N\'u{\~n}es}},
  \bibnamefont{and} \bibinfo{author}{\bibfnamefont{A.~H.}
  \bibnamefont{MacDonald}}, \bibinfo{journal}{Phys.\ Rev.\ B}
  \textbf{\bibinfo{volume}{70}}, \bibinfo{pages}{155308}
  (\bibinfo{year}{2004}).

\bibitem[{\citenamefont{Erlingsson et~al.}(20005)\citenamefont{Erlingsson,
  Schliemann, and Loss}}]{erlingsson05:035319}
\bibinfo{author}{\bibfnamefont{S.~I.} \bibnamefont{Erlingsson}},
  \bibinfo{author}{\bibfnamefont{J.}~\bibnamefont{Schliemann}},
  \bibnamefont{and} \bibinfo{author}{\bibfnamefont{D.}~\bibnamefont{Loss}},
  \bibinfo{journal}{Phys.\ Rev.\ B} \textbf{\bibinfo{volume}{71}},
  \bibinfo{pages}{035319} (\bibinfo{year}{20005}).

\bibitem[{\citenamefont{Blanter and B\"{u}ttiker}(2000)}]{blanter00:1}
\bibinfo{author}{\bibfnamefont{Y.~M.} \bibnamefont{Blanter}} \bibnamefont{and}
  \bibinfo{author}{\bibfnamefont{M.}~\bibnamefont{B\"{u}ttiker}},
  \bibinfo{journal}{Phy.\ Rep.} \textbf{\bibinfo{volume}{336}}
  (\bibinfo{year}{2000}).

\bibitem[{\citenamefont{Brataas et~al.}(2000)\citenamefont{Brataas, Nazarov,
  and Bauer}}]{brataas00:2481}
\bibinfo{author}{\bibfnamefont{A.}~\bibnamefont{Brataas}},
  \bibinfo{author}{\bibfnamefont{Y.~V.} \bibnamefont{Nazarov}},
  \bibnamefont{and} \bibinfo{author}{\bibfnamefont{G.~E.~W.}
  \bibnamefont{Bauer}}, \bibinfo{journal}{Phys.\ Rev.\ Lett.}
  \textbf{\bibinfo{volume}{84}}, \bibinfo{pages}{2481} (\bibinfo{year}{2000}).

\bibitem[{\citenamefont{Martin and Landauer}(1992)}]{martin92:1742}
\bibinfo{author}{\bibfnamefont{T.}~\bibnamefont{Martin}} \bibnamefont{and}
  \bibinfo{author}{\bibfnamefont{R.}~\bibnamefont{Landauer}},
  \bibinfo{journal}{Phys.\ Rev.\ B} \textbf{\bibinfo{volume}{45}},
  \bibinfo{pages}{1742} (\bibinfo{year}{1992}).

\bibitem[{\citenamefont{Mott}(1929)}]{mott29:425}
\bibinfo{author}{\bibfnamefont{N.~F.} \bibnamefont{Mott}},
  \bibinfo{journal}{Proc.\ R.\ Soc.\ London A} \textbf{\bibinfo{volume}{124}},
  \bibinfo{pages}{425} (\bibinfo{year}{1929}).

\bibitem[{\citenamefont{Shepard et~al.}(1992)\citenamefont{Shepard, Roukes, and
  van~der Gaag}}]{shepard92:9648}
\bibinfo{author}{\bibfnamefont{K.~L.} \bibnamefont{Shepard}},
  \bibinfo{author}{\bibfnamefont{M.}~\bibnamefont{Roukes}}, \bibnamefont{and}
  \bibinfo{author}{\bibfnamefont{B.~P.} \bibnamefont{van~der Gaag}},
  \bibinfo{journal}{Phys.\ Rev.\ B} \textbf{\bibinfo{volume}{46}},
  \bibinfo{pages}{9648} (\bibinfo{year}{1992}).

\bibitem[{\citenamefont{Lesovik}(1989)}]{lesovik89:592}
\bibinfo{author}{\bibfnamefont{G.~B.} \bibnamefont{Lesovik}},
  \bibinfo{journal}{JETP Lett.} \textbf{\bibinfo{volume}{49}},
  \bibinfo{pages}{592} (\bibinfo{year}{1989}).

\bibitem[{\citenamefont{Molenkamp et~al.}(2001)\citenamefont{Molenkamp,
  Schmidt, and Bauer}}]{molenkamp01:121202R}
\bibinfo{author}{\bibfnamefont{L.~W.} \bibnamefont{Molenkamp}},
  \bibinfo{author}{\bibfnamefont{G.}~\bibnamefont{Schmidt}}, \bibnamefont{and}
  \bibinfo{author}{\bibfnamefont{G.~E.~W.} \bibnamefont{Bauer}},
  \bibinfo{journal}{Phys.\ Rev.\ B} \textbf{\bibinfo{volume}{64}},
  \bibinfo{pages}{121202(R)} (\bibinfo{year}{2001}).

\bibitem[{\citenamefont{{van den Brom} and {van
  Routenbeek}}(1999)}]{vandenBrom99:1526}
\bibinfo{author}{\bibfnamefont{H.~E.} \bibnamefont{{van den Brom}}}
  \bibnamefont{and} \bibinfo{author}{\bibfnamefont{J.~M.} \bibnamefont{{van
  Routenbeek}}}, \bibinfo{journal}{Phys.\ Rev.\ Lett.}
  \textbf{\bibinfo{volume}{82}}, \bibinfo{pages}{1526} (\bibinfo{year}{1999}).

\end{thebibliography}
\end{document}